\documentclass[11pt]{article}

\usepackage[margin=1in]{geometry}
\usepackage{setspace}
\usepackage[T1]{fontenc}
\usepackage[utf8]{inputenc}
\usepackage{lmodern}
\usepackage{microtype}
\usepackage{tikz}

\usepackage{amsmath,amssymb,amsthm,mathtools,bm}

\usepackage{graphicx}
\usepackage{booktabs}
\usepackage{caption}
\usepackage{subcaption}
\usepackage{float}
\usepackage{rotating} 

\usepackage{enumitem}
\usepackage{xcolor}

\usepackage[round,authoryear]{natbib}
\usepackage{hyperref}
\hypersetup{
  colorlinks=true,
  linkcolor=blue,
  citecolor=blue,
  urlcolor=blue
}
\usepackage{cleveref} 

\numberwithin{equation}{section}
\newtheorem{theorem}{Theorem}[section]
\newtheorem{proposition}[theorem]{Proposition}
\newtheorem{lemma}[theorem]{Lemma}

\theoremstyle{definition}
\newtheorem{definition}[theorem]{Definition}
\theoremstyle{remark}
\newtheorem{remark}[theorem]{Remark}

\DeclareMathOperator{\E}{\mathbb{E}}

\newcommand{\R}{\mathbb{R}}

\newcommand{\1}{\mathbf{1}}

\begin{document}

\title{Public Communication with Externalities\thanks{We would like to thank Christian Hellwig for extensive feedback and invaluable guidance throughout this project. We are also grateful to Ryan Chahrour, Edouard Challe, Jacques Cr\'{e}mer, Harry Di Pei, Jacob Gershman, Alexander Guembel, Joseph Harrington, Martin Hellwig, Philippe Jehiel, David Jimenez-Gomez, Stephan Lauermann, Shuo Liu, George Mailath, Thomas Mariotti, Eric Mengus, Stephen Morris, Guillermo Ordo\~{n}ez, Alessandro Pavan, Guillaume Plantin, Fran\c{c}ois Salani\'{e}, Jakub Steiner, Jean Tirole, Robert Ulbricht, Xavier Vives, Georg Weizs\"{a}cker, Alex Wolitzky and Takuro Yamashita for many helpful comments and discussions. Financial support from the European Research Council under the European Community's 7th Framework Programme FP7/2007-2013 grant agreement N263790 (InfoMacro) is gratefully acknowledged. All remaining errors are ours.}}

\author{Georgy Lukyanov\footnote{Ecole polytechnique, CREST, 5 av. Le Chatelier, 91120 Palaiseau, France. Email: georgy.lukyanov@polytechnique.edu}
\and
Konstantin Shamruk\footnote{Toulouse School of Economics, Manufacture des Tabacs, 21 All\'{e}e de Brienne, 31000 Toulouse Cedex, France. Email: c.shamruk@gmail.com}
\and
Tong Su\footnote{Toulouse School of Economics, Manufacture des Tabacs, 21 All\'{e}e de Brienne, 31000 Toulouse Cedex, France. Email: sutongmichael@yahoo.com}
\and
Ahmed Wakrim\footnote{Ecole polytechnique, CREST, 5 av. Le Chatelier, 91120 Palaiseau, France. Email: ahmed.wakrim@polytechnique.edu}
}

\date{\today}

\maketitle

\begin{abstract}
This paper develops a model in which a sender strategically communicates with a group of receivers whose payoffs depend on the sender's information. It is shown that aggregate payoff externalities create an endogenous conflict of interests between the sender and the receivers, rendering full information revelation, in general infeasible. We demonstrate that an exogenous bias in the sender's preferences can improve public information provision and raise welfare. Two applications of the setup are discussed.
\end{abstract}



\medskip

\noindent
\begingroup
\setlength{\fboxsep}{8pt}\setlength{\fboxrule}{0.4pt}%
\fbox{%
  \begin{minipage}{0.97\textwidth}
  \small
  \textbf{Author's accepted manuscript (postprint)} of:\\
  Lukyanov, G., Shamruk, K., Su, T., \& Wakrim, A. (2022). \emph{Public communication with externalities}. 
  \textit{Games and Economic Behavior}, 136, 177--196. 
  \textbf{Version of Record (VoR) DOI:}
  \href{https://doi.org/10.1016/j.geb.2022.09.003}{10.1016/j.geb.2022.09.003}.\\[2pt]
  \textbf{License for this manuscript:} CC BY-NC-ND 4.0 
  \href{https://creativecommons.org/licenses/by-nc-nd/4.0/}{(link)}.
  \end{minipage}%
}
\endgroup

\medskip

\section{Introduction}

In their decision making, individuals often rely on information from another party. Investors use credit ratings to evaluate their investment choices; depositors assess solvency risk of their bank; employees inquire the profitability of their company and decide whether their current position is worth their time and effort; and citizens keep a close eye on government statements during an emergency. An important factor affecting individuals' decisions is \emph{interaction within a group}. An investor may not be willing to hold the security if he believes that other investors are not going to do the same.\footnote{\citet{HeXiong2012} develop a dynamic debt run model showing that the pattern of yield spreads can be largely driven by creditors' coordination concerns, whereby each of them decides whether or not to roll over. This effect makes yield spreads excessively volatile in comparison to movements in fundamentals.} A depositor has a stronger incentive to withdraw if he knows that other depositors are withdrawing so that a bank run is on its way.\footnote{In this case the endogenous conflict of interest comes from investors withdrawing their money hence making the bank less liquid, thus harming the bank's investments indirectly, due to potential liquidity shortages with a bank that might decline some loans therefore denying potentially profitable investment opportunities.} Working synergies tend to make each employee more productive, and hence more willing to exert effort, if his colleagues are exerting effort as well.\footnote{For the optimal contracting setup with synergies, see \cite{Goldsteinetal2013}.} The federal emergency declaration, despite its purpose to promote precautionary measures, may trigger a massive panic, bringing more grief and destruction.\footnote{Let us take the example of COVID more specifically, when a government announces a state of emergency people start panicking and therefore creating massive waves of consumers going to buy food. People anticipate potential shortages of food and supplies by buying enormous quantities leaving other people without food.}

In all these cases, the actions exhibit \emph{externalities} on the rest of the population. For instance, during a bank run, the depositors do not take into account that each individual withdrawal raises the likelihood of bankruptcy, albeit by a negligible amount. The presence of such externalities calls for the corrective action of the regulatory authority.

We analyze the way in which the distortion caused by the externality affects the Sender's strategic information disclosure. The central question raised in the paper may be summarized as follows: When does it pay to make public disclosures of information? In case when each agent's action imposes an externality on the rest of the population, an additional piece of information would affect his belief. Manipulation of the agents' posteriors through communication would thus partly correct for the externality.

The next quotation provides a vivid example from the credit ratings' industry:
\begin{quote}
\ldots{}the very fact of a rating downgrade has an effect in the financial market\ldots{ }even if [investors] do not rely on ratings, they pay attention to the downgrade because they consider that other market participants may react negatively to the downgrade\ldots{ }[A] confidence crisis in ratings or massive rating downgrades may totally destabilize the financial markets\ldots{ }therefore, leading CRAs are extremely reluctant to downgrade a company's debt.\footnote{See \citet{Darbellay2013}, pp. 183-185.}
\end{quote}

This suggests that the very reason that a CRA (informed party) does not want to reveal bad news to investors (decision makers) is that the rating agency worries about the effect of a downgrading on market sentiments and the potential disruptive impact of the panic. In this case, credit rating plays a dual role: on the one hand, it gives investors a better estimate of the security's payoff; on the other hand, it may induce a harmful response by the large group of investors. More informative credit ratings seem to be welfare-enhancing through the first effect, but may well be welfare-detrimental through the second effect.

This dual-role nature of public information is quote ubiquitous and is not confined to CRAs. Another illustrative case of the mechanism we have in mind can be illustrated by the recent COVID-19 crisis. The governments and health officials of the countries that were most seriously in the first several weeks, ever before the WHO officially declared it a pandemic, faced a dilemma. If it were to downplay potential dangers, that could promote reckless and imprudent behavior and the neglect of social distancing which would eventually result in the increase in mortality rates. By contrast, an overly disturbing statement could have initiated financial market hysteria and buying frenzy.\footnote{\cite{Atligetal2020} and \cite{Bakeretal2020} record the spike in economic uncertainty induced by COVID. \cite{Garrett2020} discusses the destabilizing impact of misinformation.}

This paper builds a stylized model of strategic communication, in which a single informed party (Sender) reveals her information, a signal about an unobserved state of nature ($\theta$), to a continuum of decision makers (receivers), each of whom faces a binary action choice (``risky'' versus ``safe''). Payoffs from the safe and the risky action consist of two components: the individual component, which is equal to $\theta$ for the case of the risky action and normalized to zero for the case of the safe action; and the externality component: both payoffs depend on the aggregate number (measure) of individuals who undertake the risky action, which we denote by $A$.\footnote{In the case of the COVID pandemic for instance, we would have a low $\theta$ due to an emergent situation with less solid assets and investments. $A$ would represent the panic and the run on food; the aggregate level of risky action undertaken which are basically the mergent purchases. Finally, $r$ would be the extent to which the purchases of others affect the availability of food and supplies for others thus reflecting demand and supply elasticities.}

The objective of the Sender is represented by a weighted average of the receivers' welfare and her own payoff that may contain a \emph{bias} favoring the specific action (either the safe or the risky). Before receivers choose their actions, the Sender can send a collective report, revealing her information about the state of nature. We assume that Sender's message is \emph{public} and \emph{cannot be falsified or fabricated} (that is, the Sender has hard information). However, the Sender may choose to \emph{conceal} her information. Furthermore, the Sender chooses whether to send a report \emph{after} she gets the signal, and thus cannot commit to her information disclosure strategy.

We pin down the unique equilibrium in the continuation game following the Sender's message (see Proposition \ref{prop:contgame}). In this equilibrium, receivers adopt \emph{symmetric switching strategies}: each of them undertakes the risky action if and only if his private signal passes a given level -- the threshold which is monotonically decreasing in the public message. This implies that when the Sender reports better news, receivers behave \emph{more aggressively}, meaning that they use smaller thresholds for undertaking the risky action.

In the benchmark case where the agents' aggregate behavior does not bring any externalities, we show that a benevolent Sender (the one whose payoff places 100\% weight on the receivers' aggregate payoff) always fully reveals her information. In general, however, full information revelation is not in the self-interest of the Sender in the presence of  externalities from the aggregate risky action: conditional on her  information, the Sender would typically prefer to deviate from-truth-telling. In the equilibrium constructed in Theorem \ref{prop:fulleq}, there is an \emph{interval} of signal realizations which the Sender will wish to conceal.

The \emph{endogenous conflict of interests} between the Sender and the receivers arises due to the externalities that operate through the aggregate risky action. Subsequently we show that this discrepancy can be partially corrected by the bias in the Sender's preferences. In Section \ref{sec:welfare}, we show that the marginal change in the bias term in the appropriate direction would reduce the ``non-disclosure'' interval and raise welfare. Our welfare implications contrast with the predictions from standard communication models, in which the conflict of interests is typically welfare-detrimental.

Our model can also be cast within the organizational behavior framework. Our welfare implication resembles a conventional wisdom in business that ``nice guys finish last'', which says that organizations are usually reluctant to promote an employee who is too kind to his peers to be a leader. We microfound this conventional wisdom by showing that nice leaders, who care too much about team members' welfare may be unable to transmit enough information to them. On the contrary, the team would benefit by assigning harsh leaders who are more willing to criticize than praise, or leaders who share different objectives from the team. We defer the detailed discussion of these applications to Section \ref{sec:application}.

\subsection{Related Literature}

The canonical model of strategic information transmission was pioneered by the seminal paper of \citet{CrawfordSobel1982}. As they have shown, conflict of interests between an advisor and a decision maker inhibits efficient information transmission and reduces welfare.\footnote{An early example of the cheap-talk model with multiple audiences can be found in \citet{FarrellGibbons1989}. For the overview of the work in the field, see \citet{Farrell1995} or \citet{FarrellRabin1996}. A more recent treatment can be found in \citet{Sobel2013}.} From this perspective, our paper provides microfoundations for the conflict of interests arising as a result of externalities from aggregate behavior.

Our paper fits within the literature examining how the decision maker can take advantage of the heterogeneity in the preferences or beliefs between him and his informed advisor. The decision maker can be better off because a biased advisor may have more incentives to acquire information (\citet{CheKartik2009}), advisors with opposite biases advocate and reveal more information in total (\citet{DewatripontTirole1999}), or biased advisors may be more likely to exert efforts due to behavioral reasons (\citet{Prendergast2007}).\footnote{See also \citet{Landieretal2009}, \citet{BourjadeJullien2011} and \citet{Dziuda2011}.} We contribute to this literature by showing that the exogenous bias may be welfare-improving if it is tilted in the direction opposite to the endogenous bias, thus partially offsetting it.

Our paper also relates to a recent literature which studies the social value of public information in an environment where agents' actions exhibit strategic complementarities, including \citet{MorrisShin2002} and the papers by \citet{AngeletosPavan2004, AngeletosPavan2007}. The main intuition of their papers is that more precise public information can reduce social welfare as the coordination motive makes the agents \emph{over-weight} the public signal relative to their private signals. By contrast, in our setup, even if the benevolent Sender and the receiver were to share the same information and attach the same weight to the two signals, their interests would still diverge due to externalities.

In terms of the welfare-improving policies, \cite{CornandHeinemann2008} have suggested that rather than concealing public information or reducing its precision, the Sender may wish to \emph{restricting its degree of publicity}, thereby confining his audience to a subgroup of agents.

Although the setup that we adopt is in many respects similar to that strand of literature, our motivation is quite different. First, while in their models public information is exogenously given, we derive it from the Sender's strategic behavior. Second, we provide a distinct mechanism for efficiency loss which stems from the strategic information disclosure by the Sender.

The rest of the paper is organized as follows. Section \ref{sec:model} develops the general setup. Section \ref{sec:fulleq} takes an analytically tractable version with normally distributed priors and signals; it starts with several benchmark cases and then proceeds to equilibrium characterization for the full game. Section \ref{sec:welfare} addresses welfare implications with exogenously biased public agent; it delivers of the central result of the paper -- that the Sender's bias can improve public communication. Section \ref{sec:application} considers a number of applications and discusses some possible extensions. Section \ref{sec:conclusion} concludes.

\section{\label{sec:model}The Model}

\subsection{Environment}

Consider the game played by a Sender (he) and a uniform continuum of risk-neutral receivers (she), indexed by $i\in[0,1]$. Each receiver them faces a binary action choice: a \emph{safe} ($a_i=0$) or a \emph{risky} ($a_i=1$) action. The payoff to the safe action is normalized to 0, whereas the payoff to the risky action is given by an economic fundamental $\theta\in\R$. In addition, both the payoff to the safe and the risky action are affected by the aggregate risky action $A\equiv\int_0^1 a_i  d i$. Therefore, receiver $i$'s realized payoff in state $\theta$ is given by

\begin{equation}
	\label{eq:receiver_ex_post_payoff}
	\widetilde{\pi}_i(a_i;\theta, A) = rA+a_i\theta
\end{equation}

The parameter $r$ captures the externality. If $r>0$, the there is a \emph{positive} externality (as in the example of the liquidity shortage); $r<0$ implies that there is a \emph{negative} externality (as in the COVID pandemic example). The fundamental $\theta$ is chosen by Nature and remains unobservable at the time when receivers take their actions.

Before deciding on $a_i$, each receiver $i$ observes a noisy private signal $x_i\in\R$, whose conditional distribution is normal with mean $\theta$ and precision $\beta$:

\begin{equation*}
	x_i|\theta\sim N(\theta,\beta^{-1}).
\end{equation*}

We assume that these signals are conditionally i.i.d. across receivers.

The Sender \emph{might} observe some information about $\theta$ and transmit the public message to the receivers before they undertake their actions. Specifically, we assume that with exogenous probability $p\in(0,1)$, he obtains an informative signal $y\in\R$, drawn from a normal distribution conditional on $\theta$:\footnote{The assumption that $p<1$ is crucial for our equilibrium construction. When $p=1$, equilibrium will always feature full disclosure by the ``unraveling result''; see \cite{Milgrom1981}.}

\begin{equation*}
y|\theta\sim N(\theta,\alpha^{-1})
\end{equation*}
with precision $\alpha$. With the complementary probability, $1-p$, he gets no signal. We denote this case by $y=\varnothing$. Conditional on $\theta$, the Sender's signal is assumed to be independent from all $x_i$'s.

At the start of the game, all the agents hold a normally distributed prior belief on $\theta$, with mean $\mu$ and precision $\gamma$:
\begin{equation*}
\theta\sim N(\mu,\gamma^{-1}).
\end{equation*}

\begin{remark}
For notational simplicity in our exposition, we will denote all the probability densities by $f(\cdot)$, relying on the arguments in the parentheses to convey the precise context.

	Thus we shall write 
	\begin{itemize}
		\item $f(x,y,\theta)$ to denote the probability density of the joint distribution of $x,y,\theta$;
		\item $f(y),f(x),f(\theta)$ to denote the unconditional probability density functions of $y,x,\theta$, where e.g. the marginal distribution of $x$ is given by
		$$f(x) = \int_{\Theta}\int_\R f(x,y,\theta)  d y  d \theta;$$
		\item $f(x|\theta)$, $f(y|\theta)$ to denote the probability density functions of $x,y$ conditional on the fundamental $\theta$;
		\item $f(\theta|x)$, $f(\theta|y)$, $f(\theta|x,y)$ to denote the probability density of the belief about $\theta$ given, resp., the observation of signals $x$, $y$ or both $(x,y)$;
		\item $f(x|y)$ to denote the probability density of the Sender's belief about the signal received by a receiver, if the former observes $y$;
		\item $f(y|x)$ to denote the probability density of a receiver's belief about the value of the public signal when she observes $x$.
	\end{itemize}

	In the same vein all the cumulative density functions will be denoted by $F(\cdot)$ with the expression inside the parentheses conveying the exact meaning.

	Under the normality assumptions, the functional form of all the above objects is known up to two parameters, mean and precision (inverse of variance). Along the same lines, the precisions of the random variables will be denoted by $h$ with the corresponding subscripts and means will be denoted by bars:

	\begin{align*}
	 	& h_{\theta|y} = \alpha + \gamma,& &\quad \bar \theta (y) = \int_\Theta \theta f(\theta|y)  d \theta = \frac{\alpha y + \gamma \mu}{h_{\theta|y}}\\ 
	 	& h_{\theta|x,y} = \alpha + \beta + \gamma,& &\quad \bar \theta(x,y) = \int_\Theta \theta f(\theta|x,y)  d \theta = \frac{\alpha y + \beta x + \gamma \mu}{h_{\theta|x,y}}\\ 
	 	& h_{y|x} = \frac{1}{(\beta + \gamma)^{-1} + \alpha^{-1}},& &\quad \bar y(x) = \int_\R y f(y|x)  d y = \frac{\beta x + \gamma \mu}{h_{\theta|x}} \quad \bigg( = \bar \theta(x) \bigg)\\
	 	& h_{x} = \frac{1}{\beta^{-1} + \gamma^{-1}},& &\quad \bar x = \int_\R x f(x)  d x = \mu.
	 \end{align*}

\end{remark}

We assume that the Sender is benevolent -- that is, he about the receivers' aggregate welfare. For any action profile $\{a_i\}_{i=0}^1$, his realized payoff is given by the sum of receivers' payoffs:
\begin{equation}
\label{eq:Sender_interim_payoff}
\widetilde{\Pi}\left(\{a_i\}_{i=0}^1,\theta\right)=\int_0^1\widetilde{\pi}_i(a_i;\theta,A)di.
\end{equation}

Before the receivers choose their actions, the Sender can send a public message $m$ revealing $y$. We assume that $y$ is hard information and hence cannot be falsified.

However, the Sender can voluntarily choose to withhold his information. Therefore, his reporting strategy can be equivalently described by partitioning the public message space $\R$ into the \emph{disclosure} and \emph{non-disclosure} regions. The latter would be denoted by $\mathcal{Y}^{\text{N}}\subseteq\R$: in equilibrium, the Sender will send $m=\varnothing$ whenever he receives signal $y\in\mathcal{Y}^{\text{N}}$. Otherwise, he will report truthfully and send $m=y$. Correspondingly, the message set would be denoted by $\mathcal{M}=\R\cup\{\varnothing\}$. The Sender who failed to get the signal (an event that occurs with probability $1-p$) has no choice but to send $m=\varnothing$. Hence, his behavior is non-strategic.

Another assumption we make is that the Sender \emph{cannot commit} to his information disclosure policy. Hence, the choice of $\mathcal{Y}^{\text{N}}$ has to satisfy his \emph{ex post} incentive compatibility constraints (to be specified below).

Receiver $i$ chooses her action upon observing private signal $x_i$ and the message $m$. Her strategy is a mapping
\begin{equation}
	\label{eq:actionrule}
	a_i:\R\times\mathcal{M}\rightarrow[0,1],
\end{equation}
which for any pair $(x_i,m)$ tells her which action to undertake.

The timing of the game is summarized below:

\begin{enumerate}
	\item The Nature draws $\theta$ from $N(\mu,\gamma^{-1})$;
	\item The Sender receives $y\in\R$ (with probability $p$) or $y=\varnothing$ (with probability $1-p$), while each receiver $i$ gets $x_i\in\R$;
	\item The Sender sends a message, $m\in\mathcal{M}$;
	\item Each receiver $i$ chooses $a_i$ upon observing $(x_i,m)$;
	\item Payoffs are realized.
\end{enumerate}

\subsection{Equilibrium Definition}

Let us specify the Sender's and the receivers' objective functions. Given $(x_i,m)$, receiver $i$ chooses $a_i$ to maximize her expected payoff,
\begin{equation}
\label{eq:receiver_ex_ante_payoff}
\E\left[\widetilde{\pi}_i(a_i;\theta,A)|x_i,m\right]=rA+a_i\E[\theta|x_i,m].
\end{equation}

Notice that from the perspective of receiver $i$, both the fundamental $\theta$ and the aggregate action $A$ are considered random.

Given Sender's message, we characterize the unique symmetric equilibrium for receivers in the continuation game, in which all receivers adopt the the same threshold strategy, whereby receiver $i$ chooses $a_i=1$ if and only if her private signal $x_i$ exceeds a threshold $\hat{x}(m)$; otherwise, she chooses $a_i=0$. That is,
\begin{equation}
a_i(x_i,m)=\1_{x_i\geq \hat{x}(m)}.
\end{equation}

Therefore, given the state $\theta$ and the threshold $\hat{x}(m)$, by the Law of Large Numbers, aggregate risky action $A$ would be given by
\begin{equation}
\label{eq:aggregate_action}
A(\theta,m)=\int_0^1a_i(x_i,m)d i=\int_{x\geq\hat{x}(m)}dF(x|\theta)=1 - F(\hat{x}(m)|\theta).
\end{equation}

Given this threshold strategy, the receiver whose signal realization is exactly equal to the threshold has to be indifferent between undertaking $a_i=1$ or $a_i=0$. That is,
\begin{equation}
\E[\theta|x_i=\hat{x}(m),m]=0.
\end{equation}

Correspondingly, all the receivers who get $x_i>\hat{x}(m)$ strictly prefer $a_i=1$ while all those who get $x_i<\hat{x}(m)$ prefer $a_i=0$.

Given the threshold $\hat{x}$, receivers' aggregate payoff in state $\theta$ (which is also the benevolent Sender's \emph{ex post} payoff) is given by
\begin{align}
	\label{eq:aggregate_receiver_ex_post_payoff}
	\widetilde{\Pi}(\hat{x},\theta)&=r A(\theta,m) + \theta A(\theta,m) \notag\\
	&=(r+\theta)\left(1-F(\hat x(m)|\theta)\right).
\end{align}

The Sender evaluates the expected value of social welfare \eqref{eq:aggregate_receiver_ex_post_payoff}, conditionally on his information $y$ and anticipating the receivers' response, as given by $\{a_i(x_i,m)\}_{i=0}^1$. The Sender's strategy is to choose $m$ so as to maximize
\begin{align}
\notag
	\Pi(\hat{x}(m),y)&= \int_{\Theta}\widetilde{\Pi}(\hat{x}(m),\theta) dF(\theta|y) \\ 
	\label{eq:sender_interim_payoff}
	&=\int_{\Theta}(r+\theta)\left(1 - F(\hat x(m)|\theta) \right)  dF(\theta|y).
\end{align}

Since the only choice the Sender makes concerns the disclosure of his information, given aggregate participation \eqref{eq:aggregate_action}, we can rewrite the Sender's problem as

\begin{equation}
	\label{eq:Sender_program}
	\max_{m\in\{y,\varnothing\}}\Pi(\hat{x}(m),y),
\end{equation}
subject to the constraint that for each $y$, his disclosure choice is incentive compatible.

At this point, we are ready to give the formal definition of equilibrium for our game.

\begin{definition}
	\label{def:equilibrium}
	A Perfect Bayesian Equilibrium consists of (i) a decision threshold for the receivers, $\hat{x}(m)$, (ii) an information revelation strategy for the Sender, $\mathcal{Y}^{\text{N}}\subseteq\R$, (iii) a conditional posterior p.d.f. $f(\theta|m,x_i)$ and, (iv) a measure of aggregate participation $A(\theta,m)$, such that:
	\begin{enumerate}
		\item Given $\mathcal{Y}^{\text{N}}$, for each $m\in\mathcal{M}$ a receiver's posterior p.d.f. $f(\theta|m, x_i)$ is consistent with Bayes' rule:
		\begin{equation}
			\label{bayesupdate}
			f(\theta|m,x_i)=
				\begin{dcases}
				f(\theta|y,x_i),&\text{if } m=y\\
				\left(1 - p + p \int_{y\in\mathcal{Y}^\text{N}} \frac{f(y|\theta,x_i)}{\int_{\mathcal{Y}^\text{N}} f(y|x_i) d y } d y\right) f(\theta|x_i),&\text{if }m=\varnothing.
				\end{dcases}
		\end{equation}
		\item Given $A(\theta,m)$ and $f(\theta|m,x_i)$, the decision threshold $\hat{x}(m)$ constitutes an equilibrium in the receivers' continuation game.
		\item Given $\theta\in\Theta$ and $m\in\mathcal{M}$, the aggregate participation $A(\theta,m)$ is determined by \eqref{eq:aggregate_action}.
		\item Given the receivers' action profile $\{a_i(x_i,m)\}_{i\in[0,1]}$, the Sender optimally chooses his non-disclosure region so as to maximize \eqref{eq:Sender_program}: $\forall y\in\R$,
		\begin{equation}
			\label{eq:concealment-region}
			\Pi(\hat{x}(\varnothing),y)\geq\Pi(\hat{x}(y),y)\quad\Longleftrightarrow\quad y\in\mathcal{Y}^{\text{N}}.
		\end{equation}
	\end{enumerate}
\end{definition}

\section{\label{sec:fulleq}Public Communication}

In this section, we start our analysis by characterizing the equilibrium in the continuation game played by receivers upon receiving the Sender's message, $m$ (Proposition \ref{prop:contgame}). Given our assumption that $m$ represents hard information, there can be two possibilities: either the Sender discloses the public signal, in which case we have $m=y$, or else he conceals information, which corresponds to ``silence'', $m=\varnothing$.

Then we proceed to analyzing the informed Sender's preference for aggregate action and derive his optimal strategy during the communication stage. Proposition \ref{prop:endogconflictofinterest} shows that the Sender would prefer the receivers to play the threshold that is given by $x^{*}(y)$, which is, in general, different from the threshold $\hat{x}(y)$ that the receivers use in equilibrium, whenever $r\neq0$. The magnitude of the conflict of interests between the Sender and the receivers, as given by the absolute value of the difference between those thresholds, is proportional to the size of the externalities, $|r|$.

The key result is stated in Lemma \ref{lemm:interval}, which establishes the Sender's optimal response to the strategy played by receivers upon hearing the ``empty message''. Given any threshold $\hat{x}(\varnothing)$ that the receivers might use upon hearing $m=\varnothing$, it shows that the Sender would prefer not to disclose his signal whenever it is not too high and not too low, so that the non-disclosure set constitutes an \emph{interval}. In the limit as $|r|\to0$, this interval collapses to a single point.

The section concludes Theorem \ref{prop:fulleq}, which provides the characterization of the unique equilibrium in the entire game as the fixed point between (i) the optimal receivers' threshold $\hat{x}(\varnothing)$, which is determined by the indifference condition of the marginal receiver, for a given non-disclosure region $\mathcal{Y}^N$, and (ii) the bounds of the non-disclosure set $\mathcal{Y}^N$ for a given $\hat{x}(\varnothing)$, as pinned down by the informed Sender's indifference conditions.

These results would pave the way to discussing the welfare properties of equilibrium in the next section.

\subsection{\label{sec:contgameeq}The Continuation Game}

The next proposition establishes the existence and uniqueness of equilibrium in the continuation game.

\begin{proposition}
  \label{prop:contgame}
  For any $\mathcal{Y}^{\text{N}}$, there exists a unique equilibrium of the continuation game.
  
  This equilibrium has the following properties:
 \begin{enumerate}
\item The receiver $i$ undertakes $a_i=1$ if and only if her private signal $x_i$ exceeds a threshold $\hat{x}(m)$, where $m\in\mathcal{Y}\setminus\mathcal{Y}^{\text{N}}\cup\{\varnothing\}$;
\item Upon receiving the message $m=y$, the threshold $\hat{x}(y)$ is linear and strictly decreasing in $y$:
  \begin{equation}
    \label{eq:receiver_threshold_full_info}
    \hat{x}(y)= - \frac{1}{\beta}(\alpha y + \gamma \mu);
  \end{equation}
\item If no message is transmitted, $m=\varnothing$, the threshold $x(\varnothing)$ is implicitly defined by
  \begin{equation}
    \label{eq:receiver_threshold_no_info}
    (1-p)\bar\theta(x(\varnothing)) + p \int_{\mathcal{Y}^N} \frac{\int_\Theta \theta f(y|\theta) f(\theta|x(\varnothing))  d\theta}{\int_{\mathcal{Y}^N} f(y|x(\varnothing)) d y} d y = 0.
  \end{equation}
\end{enumerate}
\end{proposition}

\begin{proof}
  See Appendix \ref{proof1}.
\end{proof}

The first part of Proposition \ref{prop:contgame} describes the receiver $i$'s optimal response given her private signal, $x_i$. Since the expected payoff from the risky action ($a_i=1$) is strictly increasing in $x_i$, the proposition states that the receiver will optimally choose to undertake it whenever $x_i$ is larger than a given threshold $\hat{x}(m)$, which may depend on the Sender's message $m$.

The second part of Proposition \ref{prop:contgame} states a relationship between receivers' equilibrium action and the disclosed public message. When the Sender reveals better news (higher $y$), receivers will collectively act more aggressively by choosing the risky action for a larger range of private signal realizations (adopting a lower switching threshold $\hat{x}(y)$). Notice, in particular, that in the limit as the Sender becomes very optimistic ($y\to+\infty$), all receivers will choose risky action ($\hat{x}(+\infty)=-\infty$), and \emph{vise versa} for extremely pessimistic Sender.

The third part of the proposition pins down the receivers' response upon hearing no information from the Sender, $m=\varnothing$, invoking the marginal receiver's indifference condition and using the Bayes' rule.

\subsection{Non-Disclosure Region and Endogenous Conflict of Interests}

We now formally introduce the Sender-optimal threshold and the endogenous conflict of interests. 
Let us define

\begin{equation*}
  x^{*}(y)\in\underset{x}{\arg\max}\ \Pi(x,y)
\end{equation*}
as the \emph{Sender-optimal threshold}, representing the receivers' optimal response from Sender's point of view, conditional on his information $y$. 

We refer \emph{endogenous conflict of interests} as the difference between the equilibrium threshold and the Sender-optimal threshold, namely 

\begin{equation}
  \label{eq:conflictofinterest}
  \Delta\triangleq\hat{x}(y)-x^{*}(y).
\end{equation}

As the aggregate action \eqref{eq:aggregate_action} is decreasing in the action threshold, the Sender prefers more action than in full-disclosure market equilibrium if $\Delta > 0$ and prefers less in the opposite case $\Delta < 0$

The next proposition establishes the general characterization of $x^{*}(y)$ and $\Delta$.

\begin{proposition}
  \label{prop:endogconflictofinterest}
  For any $y\in\mathcal{Y}$, the Sender-optimal threshold $x^{*}(y)$ is unique and given by
  \begin{equation}
    \label{eq:dictatorial_threshold}
    x^{*}(y)=-\frac{1}{\beta}\left(\alpha y+\gamma\mu+h_{\theta|x,y}r\right).
  \end{equation}

  Moreover, the endogenous conflict of interests $\Delta$ is given by
  \begin{equation}
    \Delta=\frac{h_{\theta|x,y}}{\beta} r
  \end{equation}
\end{proposition}

\begin{proof}
  See Appendix \ref{proof:endogconflictofinterest}.
\end{proof}

Proposition \ref{prop:endogconflictofinterest} shows that the magnitude of endogenous conflict of interests increases in the magnitude of externality $r$. 
furthermore, the magnitude of conflict of interests is increasing in the precision of the signal of the Sender, $\alpha$, and (iii) decreasing in the precision of the receivers' signals, $\beta$.  
Intuitively, the Sender finds the aggregate participation insufficient whenever $r>0$, and thus will choose to conceal the public signal as long as this allows to boost participation. 
However, a greater share of information, $ \frac{\beta}{\alpha + \beta + \gamma}$, ``endowed'' to the receivers may compensate for their inability to account for externality, reducing the magnitude of conflict of interest.

\subsection{Equilibrium characterization}

We now look at the first stage of our game, where the Sender strategically discloses his information, anticipating receivers' reaction as described in Proposition \ref{prop:contgame}.

In the full game equilibrium, the Sender's (non-)disclosure strategy has to satisfy incentive compatibility conditions for each type (i.e. each signal $y$ received). 
In other words, given his non-disclosure region $\mathcal{Y}^{\text{N}}$ and corresponding threshold $\hat{x}(\varnothing)$, for all $y\in\mathcal{Y}\setminus\mathcal{Y}^{\text{N}}$ we should have:

\begin{equation}
  \Pi(\hat{x}(y),y)\geq\Pi(\hat{x}(\varnothing),y).
\end{equation}

\begin{lemma}
  \label{lemm:interval}
  When $r\neq0$, in any equilibrium, a non-disclosure region is an interval. i.e. there exist $y_1,y_2$ such that
  \begin{equation*}
  \mathcal{Y}^{\text{N}}=[y_1,y_2],
  \end{equation*}
  and the optimal communication strategy of the Sender is
  \begin{equation}
    \label{eq:optimal_communication}
    m(y) = \begin{cases}
      \varnothing \quad &\text{if } y \in (y_1, y_2)\\
      y \quad &\text{otherwise. }
    \end{cases}
  \end{equation}
\end{lemma}

\begin{proof}
  See Appendix \ref{proof:lemmainterval}.
\end{proof}

The boundaries of the equilibrium non-disclosure interval are determined by the Sender's two indifference conditions. The length of the non-disclosure region is positively related to the degree of externalities, $|r|$. The stronger is this concern, the less information a benevolent Sender is willing to transmit to receivers.

As we have established in the proof of Lemma \ref{lemm:interval}, for any threshold $x$ played by the receivers in the continuation game upon hearing $m=\varnothing$, there exists a unique pair $y_1(x)$ and $y_2(x)$ determining the bounds of the non-disclosure region. In turn, the bounds $y_1(x)$ and $y_2(x)$ uniquely pin down the threshold $\hat{x}(\varnothing)$ via Bayes' rule.

The next theorem states three conditions that pin down the three equilibrium objects: the bounds for the non-disclosure region, $y_1$ and $y_2$, and the receivers' threshold $\hat{x}(\varnothing)$ which they play upon hearing the message $m=\varnothing$.

\begin{theorem}
  \label{prop:fulleq}
There exists a $\overline{p}\in(0,1)$ such that, for all $p<\overline{p}$, an equilibrium of the full game exists and is characterized by the triple $\{\hat{x}(\varnothing), y_1, y_2\}$ satisfying $y_1\neq y_2$ and the following 3 conditions:
  \begin{flalign}
    \label{fuleqrecind}
    0&=\E\big[\theta\big|\hat{x}(\varnothing),y\in\{\varnothing\}\cup\mathcal{Y}^{\text{N}}\big],\\
    \label{pindowneqthresh}
    \hat{x}(\varnothing)&=\hat{x}(y_2),\\
    \label{pindownbottomy}
    \Pi(\hat{x}(\varnothing),y_1)&=\Pi(\hat{x}(y_1),y_1).
  \end{flalign}
\end{theorem}

\begin{proof}
  See Appendix \ref{proof:fulleq}.
\end{proof}

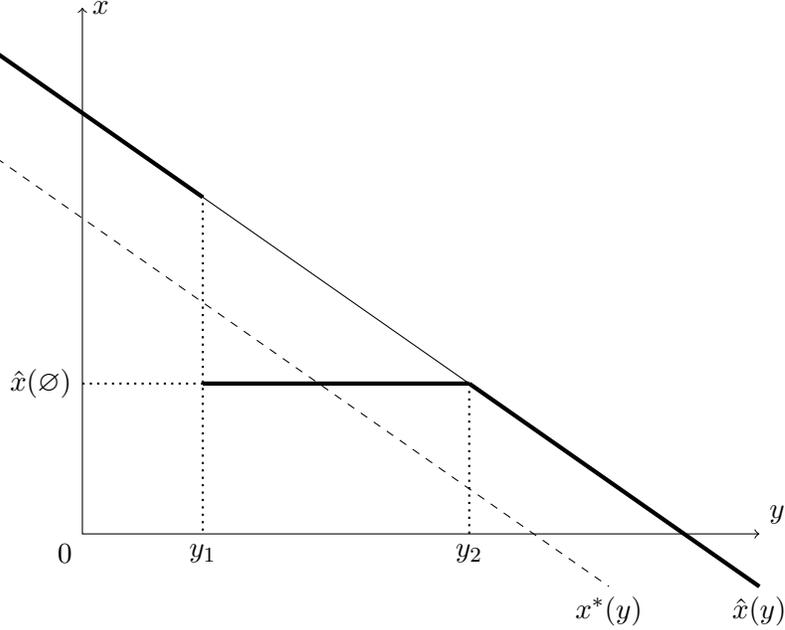
\begin{figure}[t]
  \centering
\begin{tikzpicture}[]
	\draw[<->] (-3,7) node[right]{$x$} -- (-3,0) node[below left]{0} -- (6,0) node[above right]{$y$};
	\draw[thin, dashed] (-5,5.6) -- (4,-0.7) node[below]{$x^*(y)$};
	\draw[thick, dotted] (-3,2) node[left]{$\hat{x}(\varnothing)$} -- (-1.4,2);
	\draw[thick, dotted] (-1.4,4.5) -- (-1.4,0) node[below]{$y_1$};
	\draw[thick, dotted] (2.143,2) -- (2.143,0) node[below]{$y_2$};
	
	\draw[ultra thick] (-1.4,2) -- (2.143,2);  
	\draw[ultra thick] (-5,7) -- (-1.4,4.48); 
	\draw[ultra thick] (2.143,2) -- (6,-0.7); 
	\draw[thin] (-5,7) -- (6,-0.7) node[below]{$\hat{x}(y)$}; 
\end{tikzpicture}
  \caption{Equilibrium construction $(r>0)$.}
  \label{fig:gap}
\end{figure}

Figure \ref{fig:gap} provides a graphical illustration. The solid line represents the switching threshold function $\hat{x}(y)$ used by receivers in a continuation game after the Sender's information was truthfully revealed. The dashed line represents another threshold function $x^*(y)$, referred to as \emph{Sender-optimal} threshold, which is what the Sender would pick if she can dictate the receivers' threshold rules, based on the same information as receivers in continuation games. The threshold $x^*(y)$ departs from $\hat{x}(y)$ because the Sender cares about the impact that the aggregate action $A$ has on the payoff of each receiver undertaking $a_i=1$: this the externality that is captured the $rA$ term. This is the source of the \emph{endogenous} conflict of interests that is captured by $\Delta$.

As shown in Figure \ref{fig:gap}, in the full game equilibrium the Sender's non-disclosure region is an interval, whose boundaries are determined by the two indifference conditions as in Equation \eqref{pindowneqthresh} and \eqref{pindownbottomy}. On one boundary ($y_2$ in Figure \ref{fig:gap}), the Sender is indifferent between revealing his signal or not since receivers will adopt the same strategy in either case. On the other boundary ($y_1$ in Figure \ref{fig:gap}), the receivers behave differently, namely adopt two different thresholds, with or without Sender's information. However, this boundary is pinned down by the indifferent condition that the Sender's utility is the same with receivers' using these two thresholds.\footnote{This makes use of the fact that for each $y\in\R$, the Sender's objective function $\Pi(x,y)$ is single-peaked in $x$, which is formally established and proven in the \hyperref[proof:fulleq]{Appendix}.}

Equilibrium in the full game is pinned down as the solution to the fixed point problem. Given receivers' reaction to an empty message, characterized by their strategy threshold $\hat{x}(\varnothing)$, the Sender optimally chooses the boundaries for the non-disclosure region, $y_1$ and $y_2$. And given the non-disclose region $\mathcal{Y}^{\text{N}}=[y_1,y_2]$, the threshold $\hat{x}(\varnothing)$ is optimally adopted by receivers, meaning that it is consistent with Bayesian updating conditional on receiving an empty message $m=\varnothing$. The thick solid line represents receivers' reactions upon different Sender's messages in equilibrium.

Except for the knife-edge case when $r=0$, full information disclosure is typically not feasible for the Sender even if he were benevolent. To see why, note that equilibrium threshold $\hat{x}(y)$ and the Sender-optimal threshold $x^{*}(y)$ do not coincide. If the Sender were to fully disclose his information, then the threshold played by receivers would be $\hat{x}(\varnothing)=\hat{x}_0$.\footnote{$\hat{x}_0$ is the equilibrium threshold that would be played, had receivers known for sure that the Sender had observed an uninformative signal. It is implicitly defined by $\E[\theta|\hat{x}_0,y=\varnothing]=0$.} There will always exist a Sender type $y$ who prefers $\hat{x}_0$ to $\hat{x}(y)$, and as a result, will voluntarily choose to withhold information. 

One feature of our model, which can be seen intuitively on Figure \ref{fig:gap}, is that the extent of non-disclosure, as captured by the length between $y_1$ and $y_2$, is positively related to the magnitude of endogenous conflict of interest, which is the gap between the solid and the dashed lines. Hence, a benevolent Sender will reveal less information when the externalities become stronger ($|r|$ increases).

Before we move on to discussing welfare implications and the effect of the Sender's preference bias, we make one comment regarding our equilibrium construction.  So far we did not specify receivers' out-equilibrium-beliefs that should be imposed to support our equilibrium in Theorem \ref{prop:fulleq}. Notice that, since $p<1$, there is always a strictly positive probability that the Sender does not get any informative signal. Hence, the only out-of-equilibrium play one can conceive is when the receivers observe that the Sender reports $y\in\mathcal{Y}^{\text{N}}$, i.e. some news she was not supposed to disclose. We will assume that when the receivers observe an out-of-equilibrium report $y$, they believe that it is indeed the Sender's true signal.\footnote{This is reasonable, given our assumption of hard information.} It can be immediately seen that such an out-of-equilibrium belief supports our constructed equilibrium.

\section{\label{sec:welfare}Sender's Biases and Receivers' Welfare}

We are now in a position to address the welfare properties of our setup. The question we are seeking to answer is whether giving a \emph{biased} intermediary a private access to the public signal $y$ can improve the receivers' \emph{ex ante} expected aggregate payoff. To this end, we compare the receivers' welfare under two circumstances: (i) the non-strategic Sender who always discloses the signal whenever he gets one, and (ii) the Sender who has a certain preference bias towards the receivers' aggregate action, $A$.

Substituting the optimal message \eqref{eq:optimal_communication} into the ex-post receiver's payoff \eqref{eq:aggregate_receiver_ex_post_payoff}, we obtain the ex-ante welfare as
\begin{equation}
	\begin{split}
		\label{eq:welfare}
		V &=\int_\Theta \Bigg(\left(1-p + p (F(y_2|\theta) - F(y_1|\theta))\right)\cdot (r+\theta)(1-F(\hat{x}(\varnothing)|\theta))\\ 
		&+ p\int_{y_1}^{y_2}(r+\theta)(1 - F(\hat x(y)|\theta))f(y|\theta) d y \Bigg)  d F(\theta).
	\end{split}
\end{equation}

In this section, we will introduce the \emph{bias} into the Sender's preferences. We allow for an additional exogenous bias stemming from the Sender's preferences. Specifically, we assume that the Sender cares about receivers' welfare, but also has her own preference towards risky action or safe action. 
For any action profile $\{a_i\}_{i=0}^1$, her realized payoff is given by the sum of receivers' payoffs and the bias term $bA$:
\begin{equation}
\widetilde{\Pi}\left(\{a_i\}_{i=0}^1,\theta;b\right)=\int_0^1\widetilde{\pi}_i(a_i;\theta,A)di+bA,
\end{equation}

Now the biased Sender's payoff, the analogue of \eqref{eq:sender_interim_payoff}, can be written as
\begin{equation}
  \Pi(x,y,b) = \int_\Theta\Big[(r+\theta) A(\theta,x) + b A(\theta,x)\Big] dF(\theta|y).
\end{equation}

$\Pi(x,y,b)$ gives the interim payoff of the informed Sender that chose to disclose her signal $y$, given that all receivers follow an arbitrary threshold rule defined by $x$.

Using Bayes rule and the definition of aggregate action \eqref{eq:aggregate_action} it is possible to obtain the following compact expression:

\begin{equation}
  \label{eq:sender_interim_payoff_arbitrary_threshold}
   \Pi(x,y,b) = \big[r+b + \bar \theta(y)\big] (1 - F(x|y)) + \frac{1}{\alpha + \gamma} f(x|y),
\end{equation} 
where the derivation is identical to the one we did in the proof of Proposition \ref{prop:endogconflictofinterest} with $r$ replaced by $r+b$.

Accordingly, the Sender-optimal threshold becomes
\begin{equation}
\label{eq:sendoptthresholdbiased}
x^{*}(y,b)=-\frac{1}{\beta}\left(\alpha y+\gamma\mu+h_{\theta|x,y}(r+b)\right).
\end{equation} 

We assume without loss of generality that $r+b>0$, so that $y_1 < y_2$. According to Proposition \ref{prop:endogconflictofinterest}, the limiting case when $r+b=0$ corresponds to the situation when $x^{*}(y)=\hat{x}(y)$, so that the conflict of interest is absent ($\Delta=0$).

A non-strategic Sender always transmits the public signal if it is observed, so that the receivers cannot make additional inference from a $m=\varnothing$ message. The equilibrium in the continuation game in this case is given by the fixed threshold $\hat x$ such that

\begin{equation*}
	\E(\theta| \hat x) = 0, \quad \hat x = -\frac{\gamma}{\beta}\mu.	
\end{equation*}

The ex-ante welfare under a non-strategic Sender is given by:

\begin{equation}
	\begin{split}
		\label{eq:welfare-full-disclosure}
		V_\text{full disclosure} &=\int_\Theta \Bigg((1-p)\cdot (r+\theta)(1-F(\hat{x}|\theta))\\ 
		&+ p\int_\mathcal{Y} (r+\theta)(1 - F(\hat x(y)|\theta))f(y|\theta) d y \Bigg)  d F(\theta).
	\end{split}
\end{equation}

The welfare can be decomposed into three terms: (i) the baseline welfare from full disclosure of information, (ii) the net gain due to receivers misinterpreting $m=\varnothing$ as Sender's deception, and (iii) the net gain (positive or negative) from withholding the information:
\begin{equation}
	\label{eq:welfare-decomposition}
	\begin{split}
		V=V_\text{Full Disclosure} &+ \underbrace{(1-p)\int_\Theta\left(r + \theta)(F(\hat x |\theta) - F(\hat x (\varnothing)|\theta)\right)  d F(\theta)}_\text{Misinterpretation gain, $\Delta V_\text{misinterpretation}$} \\&+ \underbrace{p\int_\Theta \int_{y_1}^{y_2} (r+\theta)\cdot (F(\hat{x}(\varnothing)\mid\theta) - F(\hat x(y)\mid\theta)) d F(\theta,y)}_\text{Info withholding gain, $\Delta V_\text{concealment}$}.
	\end{split}
\end{equation}

By changing the order of integration, we can decompose the gain from witholding information into the difference of two terms:
\begin{equation}
\label{eq:deltavconcealment}
\Delta V_\text{concealment} =  p \int_{y_1}^{y_2} \bigg(\underbrace{\Pi(\hat{x}(\varnothing),y) - \Pi(\hat x (y), y)}_\text{Sender's Concealment Gain} - b\underbrace{\left(F(\hat x(y) \mid y) - F(\hat{x}(\varnothing)\mid y)\right)}_\text{Aggregate Action Gain} \bigg) d  F(y)
\end{equation}

The overall welfare gain can be decomposed as the difference between the two terms: the expected gain of the Sender from concealing the signal, and the expected loss from the excessive aggregate action. 
The expected gain of the Sender is positive, since for $y\in [y_1, y_2]$, $\Pi(\hat{x}(\varnothing),y) - \Pi(\hat x (y), y) > 0$ by the definition of the non-disclosure region \eqref{eq:concealment-region}.
The aggregate action gain is also positive, since the no-information threshold $\hat{x}(\varnothing)$ lies below the equilibrium threshold $\hat{x}(y)$, and $F(\hat x(y) \mid y)>F(\hat{x}(\varnothing)\mid y)$.
Excess aggregate action translates into welfare loss if Sender has a bias for stronger aggregate action, $b>0$, and welfare gain in the opposite case, $b<0$. 

\begin{figure}
\includegraphics[scale=0.55]{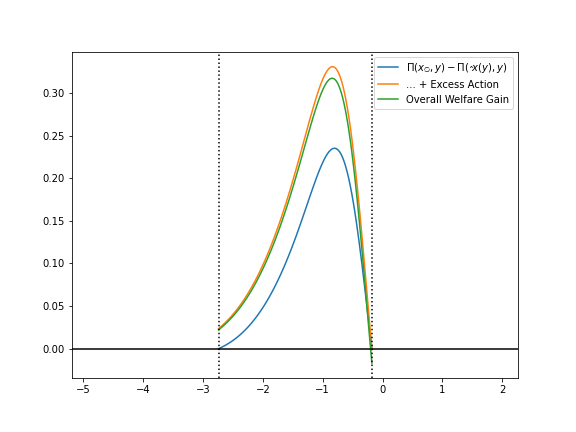}
\caption{Welfare gains from public signal concealment ($\alpha = 2, \beta = 1, \gamma = 1, \mu = 0, r = 2, b = -1/2, p = 4/5$)}
\label{fig:welfare_negative_bias}
\end{figure}

\begin{figure}
\includegraphics[scale=0.55]{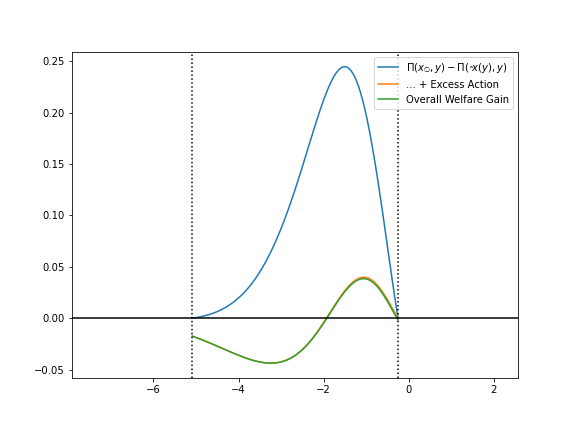}
\caption{Welfare gains from public signal concealment ($\alpha = 1, \beta = 1.5, \gamma = 1, \mu = 0, r = \frac{1}{2}, b = 1\frac{1}{2}, p = 4/5$)}
\label{fig:welfare_positive_bias}
\end{figure}

The situation for the specific set of numerical values for the parameters is represented on Figures \ref{fig:welfare_negative_bias} and \ref{fig:welfare_positive_bias}. On the horizontal axis, we plot the realization of the Sender's signal, $y$. The two dotted line enclose the concealment region: this is precisely the range of $y$'s for which the Sender's net concealment gain, $\Pi(\hat{x}(\varnothing),y) - \Pi(\hat x (y), y)$, is positive. The top Figure \ref{fig:welfare_negative_bias} shows that it may be the case that the welfare gain from concealment is everywhere positive, and so from the viewpoint of ex-ante maximization of social welfare, full disclosure clearly dominated by partial concealment. 

However, as shown on Figure \ref{fig:welfare_positive_bias}, it may also be the case that the interim welfare gain is negative for most $y$'s.

Because of the externalities, the Sender has an endogenous conflict of interests with the receivers. In turn, this suggests that the exogenous conflict of interest in the ``opposite direction'' may correct for the bias and raise the receivers' welfare. 

From \eqref{eq:sendoptthresholdbiased}, it is straightforward to see that $\frac{\partial}{\partial b}x^*(y,b)<0$. If the Sender obtains private benefit whenever more receivers undertake the risky action ($b>0$), she would like to boost participation even more compared to a benevolent Sender. Conversely, when the risky action imposes a cost to the Sender ($b<0$), she would prefer to abate participation.

Note that if $b<0$ (so that the Sender incurs a small private cost when receivers are choosing the risky action), the Sender-optimal threshold shifts up. Since it comes closer to the equilibrium threshold, the equilibrium non-disclosure region will be smaller. In other words, a biased Sender will be able to transmit more information to receivers, compared to a benevolent Sender.

This is summarized in the following proposition:

\begin{proposition}
\label{lemm:bias}
The threshold played by receivers when they get no public message, $\hat{x}(\varnothing,b)$, is strictly increasing in $b$:
\begin{equation*}
\frac{\partial\hat{x}(\varnothing,b)}{\partial b}>0.
\end{equation*}
\end{proposition}
\begin{proof}
See Appendix \ref{proof:lemmabias}.
\end{proof}

The results of the numerical simulations on Figure \ref{fig:welfare_negative_bias} illustrate that a small exogenous bias (in an appropriate direction) will increase receivers' welfare. Let us discuss the intuition for this result.

For the sake of exposition, consider the case when  $r>0$. Suppose there is a marginal decrease in Sender's bias from $b=0$ to $b=-\varepsilon$. Since the receivers' response to the information disclosed by the Sender (as given by their threshold $\hat{x}(y)$) does not depend on the Sender's bias $b$, the receivers' response to any \emph{non-empty message} $m=y\notin\mathcal{Y}^{\text{N}}$ remains unchanged.

There are two ways in which the marginal change in the Sender's bias impacts communication equilibrium (and hence $V$). First, it is known that the benevolent Sender would wish to conceal some mildly negative news -- so that the receivers are unable to distinguish the ``no news'' case with the ``bad news'' case. The marginal reduction in $b$ shifts the boundaries of the non-disclosure region: some of the $y$'s that were previously concealed and thus induced a response $\hat{x}(\varnothing,0)$ would now induce a higher response $\hat{x}(y)$ -- that is, the receivers will take a risky action less frequently for those realizations of $y$ which were not disclosed by the benevolent Sender ($b=0$) but which become disclosed under $b<0$.

Second, by Proposition \ref{lemm:bias}, the decrease in $b$ leads to a reduction in $\hat{x}(\varnothing)$: that is, now when the receivers hear no news ($m=\varnothing$) but at the same time know that the Sender is biased \emph{against the risky action}, they interpret the Sender's silence not so much to her disadvantage (as compared to the case when $b=0$). Precisely because they know that the Sender's incentives are now more aligned with their own ($x^{*}(y,b)$ gets closer to $\hat{x}(y)$ for mildly negative $b$), upon hearing $m=\varnothing$, they attribute it more to the Sender having received no information that to having received unfavorable information, and this is what induces a reduction in $\hat{x}(\varnothing,b)$ -- when the receivers obtain no news, they choose the risky action more frequently (for a larger set of private signal realizations), which benefits the Sender.\footnote{Even though for $b<0$, the Sender's preference is \emph{against} $a_i=1$, from the viewpoint of the Sender with the mildly negative $b$, the receivers are still not choosing $a_i=1$ frequently enough. (We still have $x^{*}(y,b)<\hat{x}(y)$ for most realizations of $y$).}

However, since the boundaries of the non-disclosure region, $y_1$ and $y_2$, are pinned down by the Sender's indifference conditions, the changes in $y_1$ and $y_2$ have only a \emph{second-order} (adverse) impact on the her welfare. On the other hand, the increase in $\hat{x}(\varnothing,b)$ produces a \emph{first-order} (favorable) impact on the ex-ante expected welfare $V$.

As we can see, benevolence does not lead to the highest social welfare. Similar arguments have also been pointed out by the previous papers. \citet{Rogoff1985} has shown that the society is better off appointing a central banker who does not share the social objective function but instead has a bias towards inflation rate stabilization. Another related paper is \citet{Boltonetal2013} who show that a leader with resoluteness, a bias towards her initial action choice, is optimal since she is able to better coordinate her team members, compared to a benevolent leader who constantly revises her action choices upon arrival of new information. We contribute to this literature by showing that endogenous conflict of interests arises due to the Sender's concern for correcting the externality.

Note that whether a bias is welfare-improving or not depends on the nature of externality: when receivers' risky actions bring positive externalities ($r>0$), it is better for them to deal with a \emph{conservative} Sender who is biased against revealing good news, while an \emph{aggressive} Sender, who is privately willing to reveal more good news, would improve receivers' welfare if the risky actions impose negative externalities ($r<0$). We will discuss these welfare implications in more detail for the two economic applications, credit rating agencies and leadership, in Section \ref{sec:application}. What we are certain of, however, is that whenever $r\neq0$, the absence of the bias ($b=0$) is \emph{not} optimal.

\section{\label{sec:application}Applications and Extensions}

	This section provides several economic applications that fit into the structure of our general model outlined in Section \ref{sec:model}. We discuss the positive and normative content of our key results in the context of disclosure policies by the credit rating agencies, the emergency behavior of governments, the announcements made by the central bank and the role of leadership in organizations. Then we proceed with an informal discussion of several aspects of our model. We briefly discuss out-of-equilibrium beliefs, potential sources of multiplicity and the nature of information.

\subsection{Credit Rating Agencies and Investors}

The credit rating industry as one important player in the financial market, especially the debt market, has drawn much criticism after the 2007-2008 US subprime crisis. Investors blamed rating agencies for their failure in providing accurate ratings on structured securities whose ratings tended to be ``inflated''.\footnote{Empirical studies with supporting evidence include \citet{CornaggiaCornaggia2013}, \citet{Jiangetal2012} and  \citet{BenmelechDlugosz2010}.} In addition, many have pointed out that rating inflation is a consequence from the conflict of interests between CRAs and investors since the CRAs are typically paid by issuers.\footnote{As demonstrated by \citet{Boltonetal2012}, the trusting nature of investors, due to either naiveness or regulatory constraints, makes rating inflation sustainable in equilibrium. \citet{SkretaVeldkamp2009} show that competition among rating agencies exacerbates rating inflation, even if the ratings are unbiased and independent, since the issuers can selectively announce the most favorable rating. This phenomenon that has been known as ``rating shopping''.}

There exists a number of models describing various additional strategic motives for CRAs to perform not only as passive information providers. 
First, CRAs can have real (feedback) effects on the cost of capital for issuing firms, whose financial contracts are often contingent on their credit ratings (See, e.g., \citet{KligerSarig2000}, \citet{KisgenStrahan2010}, \citet{Chenetal2012} and \citet{Manso2013}). 
As a result, a single downgrade can trigger multi-notch downgrades, known as credit-cliff, which is an undesirable outcome for both the issuer and investors. 
Second, as institutional investors are usually constrained by regulation to invest only into bonds rated above investment-grades, rating inflations can actually be in favor of those investors who can thereby earn higher yields (See \citet{Calomiris2009}, \citet{Oppetal2013} and \citet{CornaggiaCornaggia2012}). 
Third, credit ratings can act as a coordination device and be self-fulfilling. In \citet{Bootetal2006}, investors ask for high interests premium for a low-graded firm, which will optimally take the risky project, in turn, associated with higher risk.

We argue that, while all of the models cited above describe medium run interactions involving CRAs, our model complements this research by describing the short run effects of CRA behavior.
One may consider the events around an arrival of important news about the payoff of a security as a game played between a CRA, who may have information about the possible changes the fundamental, and investors, who decide whether or not to adjust their holdings of that security while the payoff from trading on the secondary market is subject to externality. 
The incentive of the CRA is exactly captured by the exogenous and endogenous conflicts of interests. 
On one hand, the CRA could be exogenously biased and willing to give issuer-favorable ratings since it usually gets side payments from the issuer. On the other hand, recall the quotation of \citet{Darbellay2013} given in the beginning of our paper, CRAs do care about the welfare of investors. So even in absence of exogenous conflict of interests the CRA, akin to the benevolent Sender in our model, is endogenously biased to give good ratings.

One popular idea among the potential CRA reforms is to replace the current issuer-pay model with investors-pay model.\footnote{See \citet{MedvedevFennell2011} for a survey on related policy proposals and debates. Also see \citet{White2010} for an excellent review on the institutional background of credit rating agencies.} For example, \citet{Diomandeetal2009} and \citet{Lynch2010} propose public-funded credit rating agencies, which ideally will represent the best interests of investors. 

We would like to argue that such attempts are worthwhile however not enough. Note that under the current issuer-pay model, the exogenous and endogenous biases of a CRA are in the same direction, giving the CRA a compounded incentive to inflate its ratings and resulting in a much inefficient communication between the CRA and investors. Yet, switching the issuer-pay model to investors-pay model does not solve the entire problem since a CRA still has an endogenous bias, as a consequence of externalities. In fact, the more a CRA cares about investors' welfare, the stronger endogenous conflict of interests will present in its objective. And this exactly prevents a investor-funded ``benevolent'' CRA from providing accurate information like what the proposers hope.

Given the externality attached to the investors' actions, the analysis in Section \ref{sec:welfare} suggests that investors would benefit if a CRA were biased against investing actions. This means that it is optimal for investors to appoint a \emph{conservative} CRA, who evaluates security payoffs more pessimistically than rational investors would do. Such conservatism gives the CRA an exogenous bias to induce less investors to choose investing action, offsetting its endogenous bias towards rating inflation, and hence benefits investors with more information revelation.

The regulator has long emphasized on their efforts to improve the transparency and the independence of CRAs\footnote{For example, see \emph{New rules on credit rating agencies (CRAs) enter into force: frequently asked questions}, European Commission memo, available from http://europa.eu/rapid/press-release\_MEMO-13-571\_en.htm.}, based on a rationale that investors are benefited from the reduction in the conflict of interests from CRAs. By pointing out that the regulator should also consider CRAs' endogenous biases, our theory also provides a rationale for the regulator to use ``asymmetric criteria'' on assessing CRAs and their ratings. In order to implement the CRAs' (optimal) conservatism, the regulator should make it more costly for CRAs release good ratings than bad ratings. For example, the regulator could adopt more strict assessment if a security is announced by CRAs as above investment-grade than below investment-grade, or could impose longer suspension for upgradings than downgradings. Also, the regulator could open more room for civil claims against CRAs, ultimately making CRAs rate securities more conservatively due to the potential civil liabilities.

\subsection{Emergency behavior of governments}

Our framework allows to gain some insights concerning governments' behavior during an epidemic. Putting aside policy measures such as announcing a nationwide lockdown (which obviously have \emph{direct} payoff effects), think of the government authorities whether to lay stress on the potential severity of the disease through extensive media coverage, decision to publish the number of new confirmed cases, etc.

Our model with the externality parameter $r<0$ may fit the following story: each citizen decides whether to observe a voluntary lockdown (``safe action'' $a_i=0$) or to continue an active economic life with little social distancing ($a_i=1$). The payoff to the latter depends on the overall severity of the disease ($\theta$, with lower $\theta$ corresponding to greater danger), but also on the number of those who choose to be active ($A$), which raises the risk of spread of the epidemic. In this case, aggregate behavior is likely to be excessively reckless, which is manifested by $\hat{x}(y)<x^{*}(y)$. The analysis in Section \ref{sec:welfare} suggests that the government that is mildly biased \emph{against} confinement ($b>0$) is more likely to disclose the the information ($y$), which would bring aggregate welfare benefits.

Alternatively, our setup allows to derive some implications for precautionary measures during the epidemic. Each receiver's choice can be interpreted as to whether one should wear a mask ($a_i=1$) or not ($a_i=0$). Correspondingly, $A$ would stand for the share of population wearing masks, whereas $r>0$ can be thought of as a marginal reduction in the likelihood of getting a disease. For example, it is known that several weeks before the WHO has labeled the COVID-19 outbreak a global pandemic, ``public officials initially discouraged masks over fears of shortages for health care providers''\footnote{See \emph{Fact check: Missing context in claim about emails, Fauci's position on masks}, available from https://www.usatoday.com/story/news/factcheck/2021/06/03/fact-check-missing-context-claim-mask-emails-fauci/7531267002/ We thank an anonymous Referee for suggesting this example to us.}.

The equilibrium non-disclosure region of our setup suggests that the public officials are likely to be more transparent regarding the effectiveness of masks ($\theta$) both when they substantially reduce the rate of spread of the disease and when they are completely useless (roughly, this corresponds to extremely high and low realizations of the public signal $y$), but at the same time the authorities would be less eager to disclose more ``mixed'' results.

\subsection{Monetary Policy and Forward Guidance}

The idea that public statements made by the Central Bank concerning its intentions can guide agents' expectations about future policy conduct has been reflected in the literature on forward guidance\footnote{Recent contributions include \citet{Campbelletal2012,Goodhart2013,Praet2013,Williams2013} and most notably \citet{Woodford2013}. Empirical studies on managing market expectations may be found in \citet{Neuenkirch2012,Neuenkirch2013}.}.

Relating to this issue, one may regard the Sender from our model as the Central Bank engaging in inflation or interest rate targeting, while the group of receivers can be thought of as firms making some strategic choice (such as the price, investment scale, entry/exit decision etc).

The literature distinguishes two forms of forward guidance: one is called \emph{Odyssean}, whereby the Central Bank commits to a particular action in the future; the other is called \emph{Delphic}, which is confined to forecasts and likely future policy actions, leaving the policymaker's hands untied. The virtue of the former is credibility and relative simplicity (cf. the Romer proposal\footnote{For details, see \citet{Romer2011}.}), however it also leaves little room for monetary discretion. The latter's main advantage is its flexibility, which comes at the expense of higher uncertainty.

Although our model does not allow the Sender to make her statements deliberately vague (e.g. by adding noise to her report), save for the extreme form of vagueness when nothing is reported ($m=\varnothing$), our results suggest that the policymaker would wish to make more precise announcements during the periods of inflationary booms or severe downturns.

In that respect, our paper contributes to the discussion of the variation in monetary policy in normal times and during crises (see \citet{Fahretal2013} for the recent survey).

\subsection{Organizations}

One conventional wisdom in business and organizations is that ``nice guys finish last''. \footnote{The original quotation was ``The nice guys are all over there, in seventh place'' by Leo Durocher, 1946.} Our model confirms this argument, showing that a benevolent leader in general is not the optimal leader. While nice people in workplace tend to show more agreeableness, sympathy and consideration for others, many research have shown that they in fact are less succsissful in their careers and are considered not suitable for being leaders (See \citet{Ngetal2005}, \citet{MuellerPlug2006} and \citet{Judgeetal2012}). Good leaders need to be able to tell people things that they do not want to hear, however, nice people often fail to do so because they care too much about others. Our model can be used to understand optimal leadership, by studying the trade-off between a leader's benevolence for her team and her willingness to reveal information.

Among the vast literature on leadership and organizations\footnote{See \citet{BoltonDewatripont2011} for a survey.}, one avenue of research studies how a leader should strategically convey her information to followers to build a more efficient team (See \citet{DewanMyatt2008}, \citet{MajumdarMukand2007} and \citet{FerreiraRezende2007}). Another avenue of research about leadership focuses on the question that why some characteristics are considered popular for leaders. For example, \citet{Boltonetal2013} explains why a leader with steadiness or resoluteness is optimal for a team. In their model, a resolute leader benefits the team by improving the coordination among her followers, since she is more willing to stick to her initial action choice, compared with a rational leader who constantly revises her action upon the arrival of new information.

Our paper joints these two avenues of research about leadership and organizations by showing that the desired characteristic of a leader should facilitate information transmission. The leader (Sender) has relevant information about the payoff from a team project while the followers (receivers) individually decide whether to exert effort. As teamwork typically requires collaboration, synergies or team morale, exerting effort by a team member tends to be an action with positive externality ($r>0$ case). As a result, a benevolent leader would have incentive to conceal, usually bad, information to prevent her team running into an unproductive equilibrium.\footnote{A leader in practice may also have other reasons to not reveal bad news, for example, if bad news destroy the confidence of her team members and hence negatively affects their performance (\citet{ComptePostlewaite2004}).} 

Though concealing bad information could be an optimal action for the leader \emph{ex post}, followers will respond to leader's strategic communication and are hardly benefited from it \emph{ex ante} due to insufficient information provision in equilibrium. Interestingly, among the famous \emph{U.S. Army's Eleven Leadership Principles}, the third principle is ``Know your soldiers and look out for their welfare'' while the fourth is ``Keep your soldiers informed''. What if there is a dilemma, as in our model, that a leader looks out for followers' welfare by exactly keeping them not informed? 

Our results show that being too benevolent as a leader is suboptimal for the team because the concern of maximizing team welfare exactly prevents herself revealing information to her followers. In fact, our model provides a rationale for organizations to choose a leader who shares a different objective with them, since leader's biases can be used as a commitment device for her to reveal more information. In particular, a desirable a leader should be privately more willing to reveal bad information to followers, namely being ``harsh''. 

\citet{Landieretal2009} share a similar conclusion as our model by showing that it can be beneficial for an organization to have a leader and team members with heterogeneous preference, though for a different driving force. In their model, the heterogeneity in preference, or the dissent, is valuable to an organization since it helps disciplining the leader to make decision more based on objective information rather than her own preference.

\subsection{Hard versus Soft Information}
In our model, the messages sent by the Sender are assumed to be hard. The game of hard information disclosure is first introduced in \citet{Milgrom1981}. Yet, the information is often modeled to be soft as well. In cheap-talk games, a message has no literal meaning and a receiver can never tell the Sender is lying or not by reading the message itself.

In the communication equilibrium analyzed in this paper, we have in mind the situations when the public information cannot be distorted or falsified, but can be concealed: the governments choosing which issues to put on the agenda during the election campaigns, or the health authorities choosing whether to release the data that is still to be confirmed by further research. 
Our setup shows that the Sender would be eager to release extreme signal realizations but prefers to conceal the signals in the intermediate range. It is an open question how our setup will be modified if the Sender could distort the signal he receives.

\subsection{\label{sec:commitment}Communication with Commitment}
In our setting, the Sender decides her message after the signal $y\in\mathcal{Y}$ has already been observed. One may ask what if the Sender were able to \emph{commit} to a revelation policy before her signals are realized, as commonly assumed in persuasion games.

In this case, it is no longer a game played by Sender and receivers, but a planning problem faced by the Sender alone. As suggested by \citet{KamenicaGentzkow2011}, in that case the Sender's maximization program could be equivalently formulated as looking for the optimal choice of a distribution of posteriors, which is Bayesian-consistent with the prior. They have shown that full revelation is optimal if and only if the Sender's objective is convex in the posteriors.

\section{\label{sec:conclusion}Conclusion}

We constructed a model to study the public communication between the sender with information and multiple decision makers who strategically interact with each other. As we have shown, when decision makers' payoffs exhibit externalities, the sender has incentive to ex post manipulate her reporting. As a consequence, an endogenous conflict of interests arises.

The endogenous conflict of interests is scaled by the magnitude of the externality. Higher $|r|$ typically results in less public communication in equilibrium, as the sender's non-disclosure region gets widen. In our model, the sender's exogenous biases in her preference can improve receivers' welfare, by offsetting the endogenous one. Receivers are better off with a conservative sender when their risky actions are strategic complements, or with an aggressive sender when actions are substitutes. 

Applying this welfare result into credit rating agency context, we conclude that switching to investors-pay model does not guarantee CRAs to provide accurate information. In fact, investors are better off by appointing conservative CRAs who are privately more willing to reveal bad information. In a leadership and organizations context, our model predicts that benevolent leader does not achieve the highest team welfare because she is not able to disclose enough information to her followers. Our results provide a rationale for organizations to select leaders who share different objectives.

\appendix

\section{Proofs}

\subsection{\label{proof1}Proof of Proposition \ref{prop:contgame}}

When receiver $i$ observes $x_i$ his expected marginal payoff from playing $a_i=1$ equals
\begin{equation}
  \pi(x_i, y)\triangleq\E\left[\theta|x_i,y\right]=\int_{-\infty}^{+\infty} \theta f(\theta|x_i,y)  d \theta.
\end{equation}

Using the normality of the signals, we obtain

\begin{equation}
  \label{eq:payoff}
    \pi(x_i,y)=\frac{\alpha y+\beta x_i + \gamma \mu}{h_{\theta|x,y}},
\end{equation}

Given reported $y$, the equilibrium threshold $\hat{x}(y)$ is implicitly defined by
\begin{equation}
  \pi(\hat{x},y)=0,
\end{equation}
or
\begin{equation}
  \label{eq:margReceiver}
  \E[\theta|\hat{x},y]=0.
\end{equation}

This could be interpreted as the indifference condition for the \emph{marginal receiver}, who receives a private signal exactly equal to the threshold, and is indifferent between choosing $a_i=0$ or $a_i=1$.

Under the normality of signals, \eqref{eq:margReceiver} can be solved for $\hat x$ to yield the equation \eqref{eq:receiver_threshold_full_info}:

\begin{equation}
\hat{x}(y)= - \frac{1}{\beta}(\alpha y + \gamma \mu)
\end{equation}

Similarly, for a given non-disclosure region $\mathcal{Y}^{\text{N}}$, we can also implicitly define receivers' threshold $\hat{x}(\varnothing)$ when they observe the empty message $m=\varnothing$:
\begin{equation}
\label{eq:marginal_receiver_indifference_no_info}
\E\left[\theta|\hat{x}(\varnothing),y\in\mathcal{Y}^{\text{N}}\cup\{\varnothing\}\right]=0.
\end{equation}

If the Sender's message is empty, $m=\varnothing$, a receiver understands that either the Sender is concealing public information or she not informed herself. Using the posterior density for this case, defined in \eqref{bayesupdate}, the \eqref{eq:marginal_receiver_indifference_no_info} gives us the following equation
 \begin{equation}
(1-p)\bar\theta(\hat{x}(\varnothing)) + p \int_{\mathcal{Y}^N} \frac{\int_\Theta \theta f(y|\theta) f(\theta|\hat{x}(\varnothing))  d \theta}{\int_{\mathcal{Y}^N} f(y|\hat{x}(\varnothing)) d y} d y = 0.
 \end{equation}

\subsection{\label{proof:endogconflictofinterest}Proof of Proposition \ref{prop:endogconflictofinterest}}

We consider the hypothetical case in which the sender directly imposes the equilibrium threshold strategy to the receivers, $x$. His optimal choice of $x$ will maximise his interim payoff, as given by \eqref{eq:sender_interim_payoff}.

Using the definitions of the Sender's ex ante payoff \ref{eq:sender_interim_payoff} and aggregate action \ref{eq:aggregate_action}, we can write down the logical chain:
\begin{align*}
	\Pi(x,y) 
	=& \int_\Theta (r + \theta) A(x,\theta) f(\theta|y)  d \theta \\
	=& \int_\Theta (r + \theta) \int_x^\infty f(\tilde x|\theta)  d \tilde x f(\theta|y)  d \theta \\
	=&\int_x^\infty \int_\Theta (r + \theta) f(\tilde x|\theta) \cdot f(\theta|y)  d \theta  d \tilde x \\
	=&\int_x^\infty \int_\Theta (r + \theta) f(\theta|\tilde x,y) \cdot f(\tilde x|y)  d \theta  d \tilde x \\
	=&\int_x^\infty \int_\Theta (r + \theta) f(\theta|\tilde x,y)  d \theta f(\tilde x|y)  d \tilde x \\
	=&\int_x^\infty (r + \bar \theta(\tilde x,y)) f(\tilde x|y)  d \tilde x \\
	=& \left(r + \frac{\alpha + \gamma}{h_{\theta|x,y}} \bar \theta(y) + \frac{\beta}{h_{\theta|x,y}} \E(\tilde x|\tilde x\geq x , y) \right) \left(1-F(x|y)\right)\\
	=& (r + \bar \theta(y)) \left( 1 - F(x|y)\right) + \frac{\beta}{h_{\theta|x,y} h_{x|y}} f(x|y) \\
	=& (r + \bar \theta(y)) \left( 1 - F(x|y)\right) + \frac{1}{h_{\theta|y}} f(x|y)
\end{align*}
where the penultimate line uses the formula for the mean of a truncated normal random variable
\begin{equation*}
\E(\tilde x|\tilde x\geq x, y) = \bar x(y) + \frac{1}{h_{x|y}} \frac{ f(x|y)}{1 - F(x|y)}.
\end{equation*}

So, given the normality of the signal, the Sender's payoff becomes
\begin{equation}
\Pi(x,y)=\left(r+\bar\theta(y)\right)(1-F(x|y))+\frac{1}{\alpha+\gamma}f(x|y).
\end{equation}

The necessary condition takes the form
\begin{equation}
\partial_x\Pi(x,y)=-(r+\bar\theta(x,y))f(x|y)=0.
\end{equation}

Evidently, since $f(x|y)>0$, the only solutions to this equation is given by
\begin{equation}
r+\bar\theta(x,y)=0,
\end{equation}
or, using the formula for $\bar\theta(x,y)$:
\begin{equation}
\frac{\alpha y + \beta x + \gamma \mu}{h_{\theta|x,y}}=-r.
\end{equation}

Rearranging terms yields \eqref{eq:dictatorial_threshold}. Using $\hat{x}(y)$ from Proposition \ref{prop:contgame}, the derivation of the conflict of interests $\Delta$ is straightforward.

\subsection{\label{proof:lemmainterval}Proof of Lemma \ref{lemm:interval}}	

The difference in the Sender's interim payoff from disclosing information versus remaining silent is given by the equation:
\begin{equation}
\Pi(\hat x(y),y)-\Pi(\hat{x}(\varnothing), y) = (r+ \bar \theta(y)) (F(\hat{x}(\varnothing)|y)-F(\hat{x}(y)|y))  + \frac{1 
}{\alpha + \gamma} (f(\hat{x}(y) |y)-f(\hat{x}(\varnothing)|y))
\end{equation}

Let us denote the function
\begin{equation}
\label{eq:defzeta}
\zeta(x,y)\triangleq\frac{f(\hat{x}(y)|y)-f(x|y)}{F(\hat{x}(y)|y)-F(x|y)}.
\end{equation}
This function is decreasing in both arguments, so that $\zeta'_x<0$ and $\zeta'_y<0$.

The Sender's payoff difference can thus be written as
\begin{equation}
\Pi(\hat x(y),y)-\Pi(\hat{x}(\varnothing), y) = \big[F(\hat{x}(\varnothing)|y)-F(\hat{x}(y)|y)\big]\left(r+ \bar \theta(y)-\frac{\zeta(\hat{x}(\varnothing),y)}{\alpha+\gamma}\right).
\end{equation}

The function $\Pi(\hat x(y),y)-\Pi(\hat{x}(\varnothing), y)$ has the following properties: (i) it is positive and tends to zero at $y\to\pm\infty$; (ii) it has three local extrema; (iii) when the Sender's signal is $y=y_2$, we have $\hat{x}(y_2)=\hat{x}(\varnothing)$, so that the payoff difference is equal to zero; (iv) when $y=y_2$ and $r\neq0$, the derivative of $\Pi(\hat x(y),y)-\Pi(\hat{x}(\varnothing), y)$ with respect to $y$ is non-zero.

Together, those properties imply that there exist exactly two values $y_1$ and $y_2$ at which $\Pi(\hat x(y),y)-\Pi(\hat{x}(\varnothing), y)=0$, and that this payoff difference is negative when $y$ lies in between those two values, so that the Sender prefers to remain silent and induce the threshold $\hat{x}(\varnothing)$ rather than reveal the signal $y$ and induce the threshold $\hat{x}(y)$.
 
 \subsection{\label{proof:fulleq}Proof of Theorem \ref{prop:fulleq}}
 
 We have to show that the system determined by equations \eqref{fuleqrecind}-\eqref{pindownbottomy} always has a solution. To do this, we collapse this system into a single equation in $x$. First, take the sub-system of two equations \eqref{pindowneqthresh}-\eqref{pindownbottomy} that for the \emph{fixed} threshold $\hat{x}(\varnothing)$, which we will refer to as simply $x$, implicitly determines the bounds $y_1(x)$ and $y_2(x)$. Lemma \ref{lemm:interval} guarantees that this system has a unique solution. We can write this system explicitly as
\begin{flalign}
x&=-\frac{1}{\beta}(\alpha y_2+\gamma\mu)\\
0&=[r+\bar{\theta}(y_1)](F(x|y_1)-F(\hat{x}(y_1)|y_1))+\frac{1}{\alpha+\gamma}[f(\hat{x}(y_1)|y_1)-f(x|y_1)]
\end{flalign}

We can rewrite the system as
\begin{flalign*}
\beta x+\alpha y_2+\gamma\mu&=0\\
(\alpha+\gamma)r+\alpha y_1+\gamma\mu-\zeta(x,y_1)&=0
\end{flalign*}

Implicitly differentiating the system with respect to $x$, $y_1$ and $y_2$, we get
\begin{equation*}
y_1'(x)=\frac{\zeta'_x}{\alpha-\zeta'_y}<0\quad\quad\quad\text{and}\quad\quad\quad y_2'(x)=-\frac{\beta}{\alpha}<0.
\end{equation*}
 
This system implicitly defines $y_1(x)$ and $y_2(x)$, which can be plugged into the first equation \eqref{fuleqrecind}, yielding
\begin{equation}
(1-p)\bar\theta(x) + p \int_{y_1(x)}^{y_2(x)} \frac{\int_\Theta \theta f(y|\theta) f(\theta|x)  d \theta}{\int_{y_1(x)}^{y_2(x)} f(y|x) d y} d y = 0,
\end{equation}
which can be rewritten as
\begin{equation}
\label{eq:combinedxrecindiff}
\int_{y_1(x)}^{y_2(x)}\left[(1-p)\bar\theta(x)+p\bar{\theta}(y,x)\right]f(y|x) d y = 0.
\end{equation}

In the limiting case when $p=0$, the expression in the integrand is independent of $y$ and is linear in $x$, since $\bar{\theta}(x)=\frac{\beta x + \gamma \mu}{\beta+\gamma}$. Furthermore, since $f(y|x)>0$, evidently, the expression on the left-hand side is positive at $x\to+\infty$ and is negative at $x\to-\infty$, and thus has at least one solution.

By continuity, the argument holds as well for $p>0$ in the neighborhood of zero.

\subsection{\label{proof:lemmabias}Proof of Lemma \ref{lemm:bias}}

First, observe that with $\hat{x}(\varnothing)$ uncganged, an increase in $b$ does not affect $y_2$. As for $y_1$, it is implicitly defined by the condition
\begin{equation*}
0=[r+b+\bar{\theta}(y_1)](F(\hat{x}(\varnothing)|y_1)-F(\hat{x}(y_1)|y_1))+\frac{1}{\alpha+\gamma}[f(\hat{x}(y_1)|y_1)-f(\hat{x}(\varnothing)|y_1)],
\end{equation*}
which is the counterpart of \eqref{pindownbottomy} for $b\neq0$, and which can be rewritten as
\begin{equation}
b=\frac{\zeta(\hat{x}(\varnothing),y_1)}{\alpha+\gamma}-r-\frac{\alpha y+\gamma\mu}{\alpha+\gamma},
\end{equation}
where $\zeta(\cdot)$ is defined by \eqref{eq:defzeta}.

The right-hand side is decreasing in $y_1$, since $\zeta'_y<0$, and so we have $\frac{\partial y_1}{\partial b}<0$ for the fixed $\hat{x}(\varnothing)$.

In turn, a reduction in $y_1$ induces an increase in $\hat{x}(\varnothing)$: an analysis of equation \eqref{eq:combinedxrecindiff} reveals that the second term in the integrand, $\bar{\theta}(y,x)$, is zero at the upper bound $y_2(x)$ and is negative for all $y<y_2(x)$. Hence, the reduction in the lower bound of the integral, $y_1(x)$, makes the whole expression on the left-hand side of \eqref{eq:combinedxrecindiff} negative, requiring an increase in $\hat{x}(\varnothing)$.

Finally, since $y_1'(x)$ and $y_2'(x)$ are both negative, an increase in $\hat{x}(\varnothing)$ induces a reduction in both bounds of the integral. The shift of the interval $[y_1,y_2]$ requires a further increase in $\hat{x}(\varnothing)$, and so on. The condition $p<\overline{p}$ that ensures equilibrium existence in Theorem \ref{prop:fulleq} guarantees that eventually, the economy will reach a new equilibrium with a higher $\hat{x}(\varnothing)$.

\bibliographystyle{apalike}
\bibliography{commu_fric}

\end{document}